# Perturbative transport experiments: to what extent do they really probe microscopic transport?

F. Sattin[1], D.F. Escande[1,2], F. Auriemma[1], G. Urso[3], D. Terranova[1]

[1]*Consorzio RFX, Euratom-ENEA Association, Padova, Italy*

[2] *Aix-Marseille Université, CNRS, PIIM, UMR 7345, 13013 Marseille, France*

[3] *CNR Istituto Per Le Applicazioni del Calcolo "M. Picone", Roma, Italy*

**Abstract**

Experiments featuring fast heat propagation, or so called "non-local" transport, were a puzzle for almost two decades. However recently it was shown, and it is recalled here, that a collective ideal MHD response of the plasma provides a quantitative agreement with these experiments, whereas transport plays just a secondary role. Then this work reviews the algebraic Matricial Approach to transport data inversion that provides a formally exact solution, as well as a quantitative assessment of error bars, limited to periodic signals. Conversely, standard transport reconstructions are shown to sometimes fail to match the exact solution. The adoption of automated global search algorithms based upon Genetic Algorithms is bound to greatly increase the probability of finding optimal solutions. Finally, the standard methods of reconstruction infer the diffusivity *D* and pinch *V* by matching experimental data against those simulated by transport codes. These methods do not warrant the validity neither of the underlying models of transport, nor of the reconstructed *D(r)* and *V(r),* even when the results look reasonable.

# 1 Introduction

Perturbative experiments are first-choice tools for investigating the dynamics of magnetized plasmas. By applying perturbations to a previous steady state, it is potentially possible to disentangle the different transport channels, and to resolve the transport matrix (Lopes Cardozo 1995). However, in the literature, scarce attention has been paid to the question whether the paradigm itself of transport used to interpret the data is really appropriate under all cases. The Convection-Diffusion Model (CDM) keeps being widely used as the most convenient framework to describe the plasma response to perturbations. However, on the one hand, the CDM is known to fail in several instances, notably in those experiments where the plasma features fast temperature propagation with sign inversion (so-called "non-local" transport). On the other hand, little is known about the accuracy of the standard method of reconstruction inferring the transport coefficients, diffusivity $D$ and pinch velocity $V$, by matching experimental data against those simulated by transport codes. This is commonly regarded as a sufficient validation, but actually it is not, as we are going to show.

In this work we make the following claims, supported with the analysis of examples taken from published literature:

(I) Experiments featuring so-called "non-local transport" are actually not probing transport. Rather, the plasma response is mostly driven by ideal MHD physics, and only secondly by transport. This stems from an MHD linear response model providing an intuitive interpretation of the phenomenology and a quantitative agreement with a large dataset of experiments. This subject is recalled in Section 2.

(II) For periodically modulated experiments, the pinch velocity $V$ and the diffusivity $D$ of the standard convection-diffusion model can be computed algebraically and pointwise by inverting a mere 2×2 matrix. This involves a controllable smoothing of experimental data and provides reliable error bars on the transport coefficients. The matrix to invert has one (possibly vanishingly) small eigenvalue in the spatial region where a source is present, which makes the uncertainty domain elongated along a $V/D$ = constant line. This method is much lighter computationally than classical transport codes. Whenever more than one harmonic can be analyzed, this method enables cross-checking its conclusions, assessing whether the convection-diffusion model is justified in its simplest version, and investigating possible extensions. This is recalled in Section 3.

(III) Transport codes look for profiles of transport coefficients within classes of predefined trial functions, which limits *a priori* the solution space. Therefore, in order to increase the reliability of such codes, it is important to use search strategies as comprehensive as possible in order to widen the solution space and to avoid sub-optimal solutions. An appealing approach dealing with both these issues is provided by Genetic Algorithms, and is developed in Section 4.

(IV) Several evidences point to a failure of the transport paradigm in some cases where the analysis by transport codes was performed. This calls into question why this conclusion had not yet been definitely reached earlier. As anticipated above, we argue that the answer is to be found in the fact that the standard strategy of transport codes amounts to replacing the precise question: "Is the estimated diffusivity/convection close to the true one ?" with a different one: "Are the reconstructed data sufficiently close to the measured ones ?". The two questions are not equivalent, and the second one admits potentially several different, but equivalent solutions (i.e., functions $D,V$). It is therefore overwhelmingly probable to find an apparently reasonable transport for many cases. We support this statement with several examples in Section 5.

## 2 MHD plasma response to perturbative experiments

*2.1 Experiments featuring "non-local transport"*

A large body of evidence in literature, addressing especially heat transport experiments performed in the past twenty years, leads to question the possibility of describing the plasma response within the framework of local transport equations, at least under some plasma regimes. This striking and long-standing evidence is provided by the intriguing response of the plasma to fast heating/cooling experiments performed in tokamaks and stellarators, under low-collisionality conditions. Anomalous results, originally observed during phases of spontaneous plasma dynamics (L-H transitions, sawteeths—see Callen and Jahns 1977, Fredrickson et al. 1990, Cordey et al. 1995), were later identified and more clearly characterized in perturbative experiments using a large variety of probes: pellets, radiofrequency, impurity injections, gas feed, current ramps (Sakamoto et al. 1991, Moret et al. 1993, Gentle et al. 1995a,b, Kissick et al. 1996, Stroth et al. 1996, Galli et al. 1999, Ryter et al. 2000, Zou et al. 2000, Hogeweij et al. 2000, Mantica et al. 2002, Andreev et al. 2002, van Milligen et al. 2002, Tamura et al. 2005, 2007, Inagaki et al. 2006, del-Castillo-Negrete et al. 2008, Sun et al. 2011, Rice et al. 2013). Their distinctive features are: (i) the extremely short time delay for the plasma reaction, even at the farthest distances from the initial perturbation, a delay incompatible with the assumptions of standard estimates for transport coefficients; (ii) the absence of any progressive damping of the electron temperature perturbation during its travel, replaced instead by (iii) a rapid sign reversal (a cooling at the edge turns into a heating at the core and *vice versa*) that takes place within a relatively narrow space interval. Evidences like these are referred to in literature as instances of *non-local transport*, to stress that the diffusive picture of transport is insufficient to accommodate them. A full account of the phenomenology is provided by Callen and Kissick (1997), Stroth (1998), and Pustovitov (2012). The first authors provide also a comprehensive view of the different theoretical scenarios envisaged.

It was acknowledged since the outset of these experiments that a single parabolic convective-diffusive equation with time-independent coefficients can hardly accommodate evidences (i) and (ii); furthermore (iii) can only be justified by invoking some energy buffer self-consistently triggered by the perturbation itself. Indeed, some success was achieved, within the diffusive framework, by letting the transport coefficients to vary suddenly in time throughout the radius (see, e.g., Gentle et al. 1995a, Kissick et al. 1996, Stroth et al. 1996, Andreev et al. 2002), which shifts the question to the nature of the mechanism driving this fast change. The position taken by some authors is that the "rigid response" is due to the triggering of some threshold instability (e.g. ITG or TEM) that switches regions of the plasma from a low-transport to a high-transport regime. Nonlinear mechanisms in plasma produce coherent localized structures with finite (and large) spatial extent that effectively short-circuit widely separated positions. There is little doubt that such structures do exist—some famous instances are streamers and zonal flows—(Doyle et al. 2007) and a recent study in LHD (Inagaki *et al* 2011) has given indications that they actually have some impact on long-range transport, too. Thus the hypothesis of locality was abandoned in these approaches, but still the ultimate drive for the plasma response was some form of microscopic transport. On the other hand, it was early acknowledged (Callen and Kissick 1997, Pustovitov 2012) that a promising candidate as carrier of the perturbation was the magnetic field: it may support fast propagation of signal via Alfvèn waves, it is tied to electron dynamics via Ohm's law and, furthermore, it is an excellent buffer of energy and momentum that can absorb from and deliver to the plasma at different locations. On the basis of these works we argue that this kind of experiment is not probing transport. Rather, the plasma response is collective, driven by MHD physics. We support this claim by presenting a full mathematical model, point out heuristically how the sought phenomenology arises out of it, and quantitatively show how it can interpret a large dataset of experiments.

*2.2 MHD model*

This model has been worked out first in Sattin and Escande (2013b), where it was also extensively compared against the whole experimental database for which time traces of the signal were published. In all cases quantitative agreement was found. The purpose of this section is to provide the basic principles of the theory, attempting to further elucidate the underlying physics. Furthermore, we address a few issues only cursorily touched in the original reference.

The basic picture of the plasma is provided within the framework of ideal magnetohydrodynamics (MHD). On top of a steady equilibrium we add two perturbative source (or sink) terms: for the particles ($S_n$) and the pressure ($S_p$). The ideal MHD equations linearized around the stable equilibrium write (Freidberg 2007, p. 310)

$$\begin{cases} \dfrac{\partial \rho}{\partial t} + \nabla \cdot \left( \rho^{(0)} \mathbf{v} + \rho \mathbf{v}^{(0)} \right) = S_n + \chi_n \nabla^2 \rho \\ \rho^{(0)} \left( \dfrac{\partial \mathbf{v}}{\partial t} + \mathbf{v} \cdot \nabla \mathbf{v}^{(0)} + \mathbf{v}^{(0)} \cdot \nabla \mathbf{v} \right) = \mathbf{J}^{(0)} \times \mathbf{B} + \mathbf{J} \times \mathbf{B}^{(0)} - \nabla p \\ \dfrac{\partial p}{\partial t} + \mathbf{v} \cdot \nabla p^{(0)} + \mathbf{v}^{(0)} \cdot \nabla p + \dfrac{\gamma p^{(0)}}{\rho^{(0)}} \left( \dfrac{\partial \rho}{\partial t} + \mathbf{v} \cdot \nabla \rho^{(0)} + \mathbf{v}^{(0)} \cdot \nabla \rho \right) = S_p + \chi_p \nabla^2 p \\ \dfrac{\partial \mathbf{B}}{\partial t} = \nabla \times \left( \mathbf{v} \times \mathbf{B}^{(0)} + \mathbf{v}^{(0)} \times \mathbf{B} \right) \\ \mu_0 \mathbf{J} = \nabla \times \mathbf{B} \end{cases} \quad (2.1)$$

The equilibrium quantities are labelled with superscript "$^{(0)}$". We consider adiabatic processes in the plasma, hence γ, the specific heat ratio, takes its value 5/3. We highlight that the particle and pressure equations (first and third line in Eq. 2.1) do contain some phenomenological diffusive terms ($\chi_n \nabla^2 \rho, \chi_p \nabla^2 p$) that account for background turbulent transport. The presence of the $\chi_{n,p}$'s is also required mathematically, not just for physical reasons: they are also sub-grid viscosities needed to stabilize numerically equations.

Equations (2.1) are strongly simplified under the following hypotheses:

(I) Cylindrical geometry and radial symmetry is assumed. This is a first-order guess that allows retaining just one spatial coordinate. Furthermore, we specialize calculations to low-β tokamak magnetic geometry.

(II) We do not consider the effect of finite density gradients: $\nabla \rho^{(0)} = 0$. Actually, in the experiments considered, normalized temperature gradients are usually larger than density ones: thus we retain the former ones only. This allows dropping some terms and making Eqns. (2.1) more manageable.

(III) We postulate an ordering such that the $\chi_{n,p}$'s drive a slower response than the other terms, i.e. their contribution is smaller than the one due to other terms appearing in the equations. This is consistent with their usage as phenomenological damping terms. Notice, however, that the opposite limit makes sense, too: in those regimes where the $\chi_{n,p}$'s dominate all other terms, the equations for $p$ and $\rho$ are decoupled from the other ones, and reduce to standard diffusive equations. Hence, by varying the relative weights between **v** and the $\chi_{n,p}$'s Eqns. (2.1) smoothly interpolate from a MHD-driven regime to a diffusion-driven one. Indeed this is actually what is expected after the external perturbations have been switched off, and the plasma returns to its equilibrium state obeying the diffusive dynamics.

(IV) One can neglect the $\partial \mathbf{v}/\partial t$ derivative in the momentum equation (the second line of 2.1) on account of small plasma inertia. Retaining this term is tantamount to tracking plasma dynamics over Alfvèn time scales, which are of no concern here. The momentum equation becomes therefore the standard force balance equation.

(V) Finally, the $\mathbf{v}^{(0)}$ terms stand for an equilibrium pinch that cannot be estimated in the present perturbative analysis, but requires a more comprehensive treatment, without linearization and retaining finite resistivity. The conclusion shows that $|\mathbf{v}^{(0)}| \ll |\mathbf{v}|$, hence in the following we will set $\mathbf{v}^{(0)} = \mathbf{0}$ altogether. This proof is straightforward, but is postponed to Appendix A, in order not to burden the main text.

It is convenient to choose the following normalizations: lengths are normalized to the minor radius *a*, masses to ion mass, magnetic field, density and temperature to their on-axis values, speeds to the Alfvèn velocity $u_A$, times to $a/u_A$. For simplicity, throughout the rest of the paper, we will suppose that the ratios between ion and electron temperature are equal to unity, both for the equilibrium part and for the perturbation. In Sattin and Escande (2013b) it is argued that only small quantitative corrections are involved when changing these ratios. In conclusion, the temperature *T* and pressure *p* are related through

$$T = \frac{p}{2} - \rho T^{(0)} \qquad (2.2)$$

We will refer to the full Eqns. (2.1) when numerical results will be needed. However, an accurate analysis of these equations can be accomplished by discarding terms of order $\beta, (Rq)^{-1} \ll 1$, where $R$ is the major radius and $q$ the safety factor. This leads to

$$\begin{cases} \dfrac{\partial \rho}{\partial t} + \dfrac{1}{r}\dfrac{\partial}{\partial r}(rv) = S_n + \chi_n \nabla^2 \rho \\ 0 = -\dfrac{\partial}{\partial r} S_p + \dfrac{\partial}{\partial r}\left[\dfrac{1}{r}\dfrac{\partial}{\partial r}(rv)\right] \\ \dfrac{\partial p}{\partial t} = S_p + \chi_p \nabla^2 p \end{cases} \qquad (2.3)$$

In particular, the second line of (2.3) yields straightforwardly

$$v(r) = -r \int_0^1 dz\, z S_p(z) + r^{-1} \int_0^r dz\, z S_p(z) \qquad (2.4)$$

More details about the explicit derivation of these equations are provided in Appendix A, second part. Notice that Eq. (2.4) is still compatible with a time-dependent $S_p$ as long as the latter varies slowly over Alfvènic time scales : then $v$ adjusts adiabatically its value to the instantaneous value of the source.

The prompt response of the plasma is thus easily understood: the velocity field $v$ is a compressible **E×B** drift, where the electric field arises in Faraday equation in order to vary the magnetic fields and the currents of the amount needed to fulfil at all times the force balance equation throughout the whole plasma; it drives the compressional dynamics of both matter and pressure.

The sign reversal of the response from edge to centre is understood by accounting for the following points that arise from detailed inspection of earlier equations:

(P1) the sum of the plasma and the magnetic pressure remains constant.

(P2) Conservation of the total magnetic flux inside the device.

(P3) The ideal MHD frozen flux hypothesis: particle trajectories track magnetic field lines.

(P4) In the absence of sources $|\partial_t p| \ll |\partial_t \rho|$.

Points (P1,P2,P3) are valid for short times, smaller than diffusive/collisional ones. In order to fix ideas, let us consider the specific case of a pure edge pressure sink: $S_n = 0, S_p(a) < 0$, and of a vanishing $S_p$ inside the plasma. Therefore, at the edge, $S_p < 0 \to \partial_t p < 0 \to \partial_t B_z > 0$ by virtue of (P1). Accordingly, (P2) forces $\partial_t B_z < 0$ in the core, and (P3) implies that there $\partial_t \rho < 0$ as well. By virtue of (P4), any variation in the density must be compensated by a variation of temperature in order to keep pressure almost unvaried, and (2.2) shows that this requires $\partial_t T > 0$. In conclusion, cooling the edge does heat the core.

The rationale of the model is the following: plasma cooling/heating is approximately an adiabatic process, in which density and pressure are driven by the velocity field *v*: the velocity field that appears in Ohm's and Faraday equation. Since *v* depends on the source terms, as long as $S_p$, $S_n$ are turned on, the plasma response is driven by inductive effects. Conversely, after switch-off of the source, *v* = 0 and only small-scale diffusive transport remains. Since *v* sets on almost simultaneously throughout the whole plasma once the source is turned on, any transport analysis abstracting from magnetic induction phenomena, would interpret it as an expression of "non-local" processes in the plasma.

*2.3 Some results and discussion*

We provide first a representative example out of the whole dataset considered by Sattin and Escande (2013b). Figure (1) displays the time traces of electron perturbation at four radial positions following a pellet injection, adapted from the figures in the original reference (Zou et al. 2000), where the whole experiment is described in detail. A shallow pellet produces a cooling in the outer plasma region (the time trace at *r* = 0.63 in Fig. 1), but already the neighbouring point, at *r* = 0.44, features the sign reversal of the temperature perturbation. The best fit of our simulation is displayed as solid curves. In these simulation the pellet has been modelled fairly roughly: its precise dynamics has not been considered, rather it has been modelled as a spatially fixed heat sink and particle source with gaussian shape centered at about *r* = 0.9 and with an amplitude fading away exponentially with an e-folding time of about 30 ms. Despite this, the overall agreement is quite good.

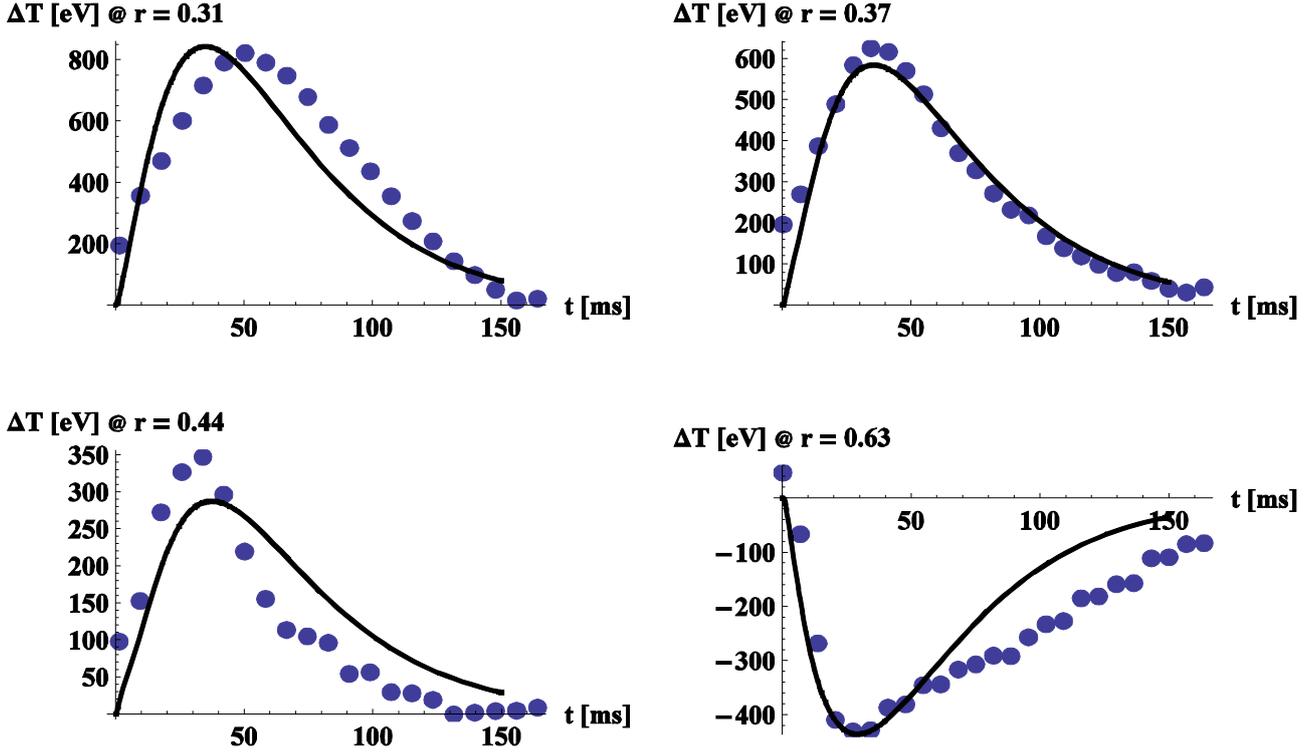

**Fig. 1**. Symbols are representative of measured temperature signals at different radial locations, and have been acquired from the original reference (Zou et al. 2000). Solid curves are the best-fit results obtained within the present model.

Although, within this level of description, the *fine* details of the sources are not relevant, this must not be regarded as a proof that the details altogether of the source are irrelevant for the modelling. On the contrary, the response driven by $v$ depends upon them, and this leads to several important corollaries. For instance, it allows explaining in a very natural way the phenomenology encountered in the JET L-mode discharge 55809 (del-Castillo-Negrete et al. 2008), which so far challenged the CDM (see discussion in Mantica et al 2008). We now describe again this case already presented in Sattin and Escande (2013b), because it is also useful for the discussion in sections 3 and 4. During this discharge two different heat transport experiments were performed: (i) the periodic modulation of central RF heating (ICRH in mode conversion scheme) allowed to assess that the electron temperature profile was likely lying on the marginal stability threshold for some instability in the range $\rho > 0.4$, and was sub-marginal in the core; accordingly, the associated transport should be high in the edge and low in the core. (ii) However, cold pulses produced by particle injection at the edge were found to travel with large velocity even in the sub-threshold region. This led the authors to claim the inconsistency of the heat transport in the two experiments. The same conclusion was later verified by other means by Sattin and Escande (2013a)—see Section 3. However, within Eqns.

(2.1-2.4) it arises just as a natural consequence, since the two sources do produce markedly different $v$ profiles. Figure (2) shows that the two different MHD responses match very well those displayed by the plasma.

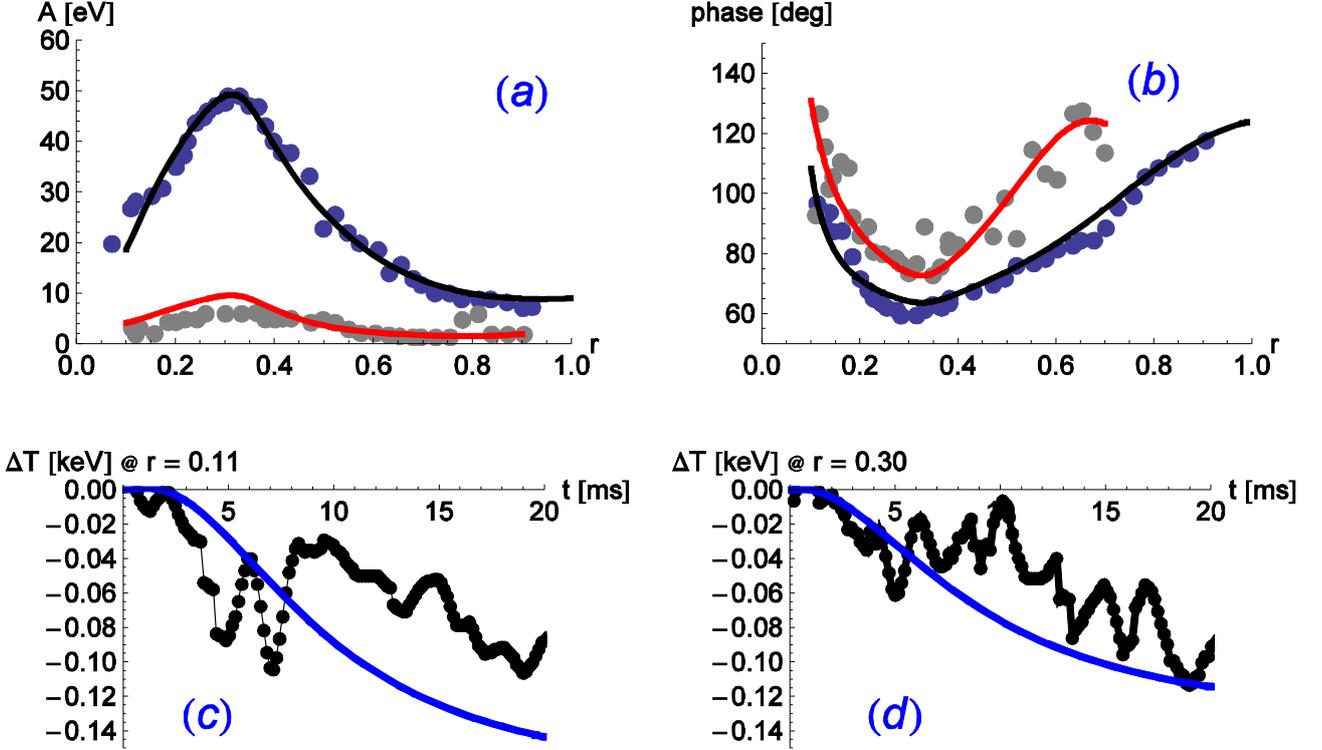

**Fig. 2**. (a) Amplitude and (b) phase of the signal from ICRH modulation for the dominant harmonic and the first odd harmonic in JET discharge 55809; (c, d) time traces of the temperature signal following the cold pulse at two radial positions. In all plots, symbols are the experimental data, solid curves, the results from our model. More details about how the simulations were carried out are to be found in (Sattin and Escande 2013b).

Turning back to the response driven by $v$, we note that the amplitude of $v$ is proportional to the external source. As a corollary, as $S_p \to 0$, there is a gradual transition from an MHD-driven "non-local" dynamics, to a standard local diffusive one. This is consistent with early findings that a minimal finite amount of power is needed to observe "non-local" phenomena (Callen and Kissick 1997). Furthermore, let us consider the stylized case of a very narrow source centred somewhere inside the plasma: $S_p = \delta(r - r_0), 0 < r_0 < 1$. Then, Eq. (2.4) yields

$$v = \begin{cases} -r_0 r & r < r_0 \\ \dfrac{r_0}{r}(1 - r^2) & r > r_0 \end{cases} \qquad (2.5)$$

That is, *v* points inwards inside the position of the source, and outwards outside, yielding an example of bidirectional transport. Examples of this kind of behaviour were found in TJ-II (van Milligen et al. 2002).

In conclusion, this model accounts—not only qualitatively but also quantitatively—for most of the so-called "non-local" phenomenology that so far escaped all attempts of understanding. It involves only six parameters to be fitted: two diffusivities and four parameters defining the sources. This is less than in a typical transport model, since we do not account for spatial variations of the diffusivities.

Some points still remain open to investigation, though. An important one is the mechanism that discriminates between ordinary diffusive transport and "non-local" one. It is acknowledged that the latter one occurs only under low-collisionality conditions (Callen and Kissick 1997). Rice et al (2013) argue that above some critical collisionality, some microinstability sets in. This increases diffusive transport, which may overwhelm the MHD component of the plasma response.

Another corollary is that any heat perturbation must involve necessarily a density one, even whenever no particle source is explicitly present. To the best of our knowledge no detailed monitoring of density behaviour has been done during non-local transport experiments. However, this is consistent with density pump-out and pump-in effects observed during RF heating in tokamaks (Angioni et al. 2004, Doyle et al. 2007, Mlynek et al. 2012, Song et al. 2012). Magnetic perturbations are expected to be produced as well, however their detection should be fairly difficult: magnetic signals are recorded using edge probes and, at the edge, the amplitude of the magnetic component of the perturbation vanishes.

A further intriguing and yet unexplained phenomenology concerns the "resurgences", namely those cases where a temperature pulse is observed to suddenly vanish and then recover back its original amplitude without reversing its sign while still travelling toward the core. To the best of our knowledge only three such instances are documented in literature: pellet experiments in LHD (del-Castillo-Negrete et al. 2010), modulated RF experiments in Asdex-Upgrade (Ryter et al. 2000) and Tore Supra (Song et al. 2012). In the latter paper no explicit identification of resurgences was made, but it can be inferred from the published results. It is not yet clear whether this kind of phenomenology can actually be enframed within a formalism like (2.1) or not. Actually, the phenomenology itself has not been well characterized yet, and the conditions leading to its appearance are not clear.

We conclude this section mentioning that "non-local transport" is not specific to magnetized plasmas. Rather, evidences for the same phenomenology have been known in condensed matter physics, namely in viscous fluids close to the glass transition (Papini 2012). Intriguingly, the theoretical interpretation worked out in that paper admits a perfect correspondence with our own.

## 3 Matricial Algorithm for extracting transport coefficients

The previous section questioned the postulate that any fast plasma response to a perturbation must necessarily be due to transport processes. Therefore it is necessary to devise tools able to infer as precisely and reliably as possible whether the data produced in some experiments are compatible with the dynamics driven by transport and — in the positive case — to give a quantitative estimate of transport itself. In this section we recall one such algorithm.

*3.1 Convective-Diffusive transport equation*

Within the standard fluid local picture the transport of the generic quantity $f$ (particle density, temperature, ...) is governed by its continuity equation

$$\frac{\partial f}{\partial t} = -\nabla \cdot \Gamma(f) + S_f \qquad (3.1)$$

where $\Gamma$ is the flux and $S_f$ accounts for sources and sink terms. Parameter $f$ stands for any of the relevant plasma parameters (particle density, electron or ion temperature, …). The dynamics of any parameter $f$ is usually affected by all the other plasma parameters ($g, h, …$), thus the flux is written: $\Gamma = \Gamma(f, g, h,...)$ or, most commonly, as a linear combination of the gradients of the quantities themselves:

$$\Gamma = -D_{ff}\nabla f - D_{fg}\nabla g + ... + V_f f \qquad (3.2)$$

The convective coefficient $V$ contains spatial derivatives of other quantities not explicitly included: in Escande and Sattin (2007, 2008) it is recalled that — even when dealing with a single quantity $f$ — a convective term $V$ must necessarily arise when the plasma is inhomogeneous. Notice that coefficients ($D_{ff}$, ...) are usually non-linear functions of the plasma parameters, hence depend on the same set ($f, g, h, ...$). From now on, we make the hypothesis that any involved perturbation is small enough to make reasonable the linearization of (3.1, 3.2) around a steady state (see Jacchia et al.

(2005) for an assessment of the contribution of non-linearities in perturbative transport). After linearization, the first-order terms in (3.1) write as

$$\frac{\partial f^{(1)}}{\partial t} = \nabla \cdot \left( \tilde{D} \nabla f^{(1)} - \tilde{V} f^{(1)} \right) + \tilde{S} \qquad (3.3)$$

In (3.3) quantities (*D,V*) are capped in order to highlight that they depend on the working plasma condition. The dependence is particularly strong in the case of stiff profiles (critical-gradient scenarios), where the flux may wildly vary in response to very small changes in the gradients (Ryter et al. 2001a,b, Mantica and Ryter 2006, Ryter et al. 2010). During perturbative experiments just one quantity *f* is customarily monitored with accuracy, whereas the other ones (*g, h*, ...) are known only approximately, or not at all. As a consequence, the $\tilde{S}$ term includes not only the variation to the source term required for the perturbative experiment to take place, but also the flux contribution due to other transport channels: i.e., the part dependent on ($g^{(1)}, \nabla g^{(1)},...$). If this last part is not negligible, therefore, $\tilde{S}$ must be regarded actually as a time-dependent and unknown quantity: the modelling stage must include $\tilde{S}$ as well among the parameters to be solved for. Failure in accounting relevant coupled transport channels may end into flawed conclusions; a couple of examples is provided in Appendix B.

*3.2 Matricial Algorithm for extracting transport coefficients*

Two classes of experiments must be distinguished on the basis of the experimental setup: modulated and pulsed ones. Perturbative experiments with periodic modulation of the input are a first-choice tool, due to their several advantages: any modulation cycle can be regarded as the repetition of the same measurement, thus enhancing the signal-to-noise ratio; furthermore, from the point of view of the analysis, they allow a remarkable simplification, since, discarding the initial transients, one can get rid of all time derivatives through a Fourier transform and reduce Eq. (3.3) to an ordinary differential equation:

$$-i\omega f = \frac{1}{r}\frac{d}{dr}\left[ r\left( D\frac{df}{dr} - V f \right) \right] + S \qquad (3.4)$$

Eq. (3.4) is written in cylindrical geometry, which is a reasonable simplification in most cases. Furthermore, the upper scripts and the caps appearing in (3.3) are now dropped for simplicity; we do not distinguish explicitly between a quantity and its Fourier transform, since the context makes clear which one is being used.

Although the original meaning of Eqns. (3.1,3.4) is to describe the dynamics of $f$, it is noteworthy that the roles can be reversed: Eq. (3.4) can be read as a complex-valued first-order ordinary differential for $D,V$, while $f$, its derivatives, and $S$ are known:

$$r\frac{df}{dr}\frac{dD}{dr} - rf\frac{dV}{dr} + \frac{d}{dr}\left(r\frac{df}{dr}\right)D - \frac{d}{dr}(rf)V = r(S + i\omega f) \tag{3.5}$$

This highlights the aspect of *inverse problem* of the computation of transport coefficients. Equations (3.4,3.5) can be solved by quadrature: by integrating Eq. (3.4) over $r$, taking into account the boundary condition $\Gamma(r=0) = 0$, one gets[1]

$$D\frac{df}{dr} - Vf = -r^{-1}\int_0^r z(S(z) + i\omega f(z))\,dz \tag{3.6}$$

Eq. (3-6) is a complex-valued algebraic equation for $(D(r), V(r))$, equivalent to a couple of real-valued ones by taking separately its real and imaginary parts. In matricial form, this couple of equations writes $\mathbf{Y} = \mathbf{M}^{-1}\cdot\mathbf{H}$, where $\mathbf{Y}$ and $\mathbf{H}$ are two-dimensional arrays: $\mathbf{Y} = (D,V)$, $\mathbf{H}$ comes from the r.h.s. of Eq. (3.6), and $\mathbf{M}$ is a 2×2 matrix. For short we refer to this algorithm as the *Matricial Algorithm* (MA); it has been known since long and rediscovered independently several times, starting from Krieger et al. (1990), then Moret et al. (1993) and Takenaga et al. (1998), up to Escande and Sattin (2012) that discussed its theoretical underpinnings. Equation (3.6) thus prescribes the *exact* couple $(D,V)$ needed to produce the measured signal $f$. It allows for an accurate estimate of error bars as well. The most convenient procedure is through a Monte Carlo approach: the data $f$ are perturbed by the addition of a random amount of noise — taken according to a statistics that mimics the expected experimental noise in terms of distribution and amplitude — and $(D,V)$ are computed accordingly. Iteration of the procedure over a large number of statistically independent runs allows to build a whole distribution of $(D,V)$ estimates, from which mean values as well as confidence intervals can be deduced [see Fig. (1) of Escande and Sattin (2012)]. Notice that, besides the error due to experimental uncertainties, there is a second source of error stemming from the interpolation scheme chosen for $f$. Indeed, though $f$ must be a smooth twice differentiable function as a solution of differential equation (3.4), operationally it is obtained as a discrete set of data $\{f_i\}$ which must be interpolated with differentiable functions prior to using Eq. (3.6). The choice of these functions yields some arbitrariness in the estimate of $df/dr$, in equation (3.6). Fortunately this is a fully verifiable procedure: the effect of different approximation schemes upon final results can be straightforwardly tested, and bears on the estimate of error bars. Throughout this work cubic splines are employed.

---

[1] A straightforward alternative to (3.6) is by integrating (3.4) from the edge $r = 1$ inward: $D\dfrac{df}{dr} - Vf = -\Gamma(1) + r^{-1}\int_r^1 z(S(z) + i\omega f(z))\,dz$. It may be convenient to avoid the source region when it is central. However, then the edge flux $\Gamma(1)$ must be either known from the experiment, or guessed.

If the frequency spectrum of the perturbation contains more than one harmonic, several replica of Eq. (3.6) can be written down, one for each frequency ω at which a response $f$ is measured. These extra constraints can be used for fixing further parameters, e.g., the details of the source. One example is given in (Escande et al. 2012, Sattin et al. 2012). Otherwise, if the number of constraints exceeds the number of parameters, the problem becomes over-determined. This over-determination can be used for consistency checks: if, within the confidence intervals set by experimental uncertainties, different estimates of the coefficients from different frequencies happen to overlap, this represents a consistency check about the validity and completeness of the model given by Eq. (3.6).

When expressing $f$ in terms of an amplitude and a phase, $f = Ae^{i\phi}$, the linear transformation that inverts Eq. (3.6) solving for $(D,V)$ becomes singular at points where $d\varphi/dr = 0$, since $\det(M) = A^2 d\varphi/dr$. This condition regularly identifies the position of the source (Escande and Sattin 2007, 2008). Hence trouble may occur when the source encompasses the spatial region of interest. This is just a consequence of the fact that the effects of transport and of the source deposition are too blended to be distinguishable. As a matter of fact, people usually try and measure transport away from the sources. However, away from the source, this algorithm works very well, as shown earlier on synthetic data (Escande and Sattin 2012), and on actual data from experiments. In particular, it enabled to model momentum and heat transport experiments, in which the source is central (Sattin et al. 2012, Escande et al. 2012, Sattin and Escande 2013a); more about this latter study is presented below.

Whenever the experimental data are good enough, the MA can check whether results of perturbative experiments are actually consistent with the diffusive paradigm (3.4). Next section discusses two such instances. In the first one the MA provides an affirmative answer to this question. The second experiment was discussed with Fig (2) in the previous section. In accordance with the conclusions of that section, the MA provides a negative answer.

*3.3 Success and failure of the simplest diffusive transport paradigm*

Modulated experiments can be carried out with a variety of actuators, but the modulation of radiofrequency heating is outstanding as far as heat transport studies are concerned: ECRH or even ICRH in mode conversion scheme (Mantica et al. 2008). It has several qualities: a narrow deposition profile, that can also be varied radially at will; it is easy to vary its duty cycle and therefore its harmonic content. Finally, since RF delivers only energy and not matter, its impact upon the density channel may be deemed as small. This works as a first-order approximation, although it is not completely true (e.g., the well known density pump-out). Furthermore, in the specific case of ICRH there is the further issue that power is delivered to both the electron and ion temperature channels (Ryter et al. 2011).

As a first exercise we consider the ECRH modulation experiments at ASDEX-Upgrade reported in Mantica et al. (2006). The source was modulated off-axis ECRH placed at about two-thirds of the radius. The off-axis position of the source was explicitly designed to investigate the existence of a pinch term in the heat transport. The modulation was a square waveform with a duty-cycle $d = 0.88$, and dominant harmonics at either 14.7 or 29.4 Hz. For the present work the very high quality of the temperature measurements is particularly interesting; it was performed with a 60 channel ECE heterodyne radiometer. The resulting profiles $\{A_i, \varphi_i\}$ appear remarkably free from noise. In this study we use specifically the data produced in shot 17175, reported in figures (4,5) of the original paper and reproduced in Fig. (3) below. We consider only the data for the first three harmonics, out of the five reported in Mantica et al. (2006): the discarded ones feature too unfavourable a signal-to-noise ratio. Data are fitted using linear combinations of even Hermite polynomials. Minor radius is $a = 0.65$ m, and electron density is held constant throughout the radius, $n_e = 2 \times 10^{19}$ m$^{-3}$. The calculation of error bars follows the guidelines given in the earlier section, and was done postulating a statistical error of 7% both on the amplitude and the phase, consistently with what reported in Mantica et al. (2006); 100 statistical runs are performed.

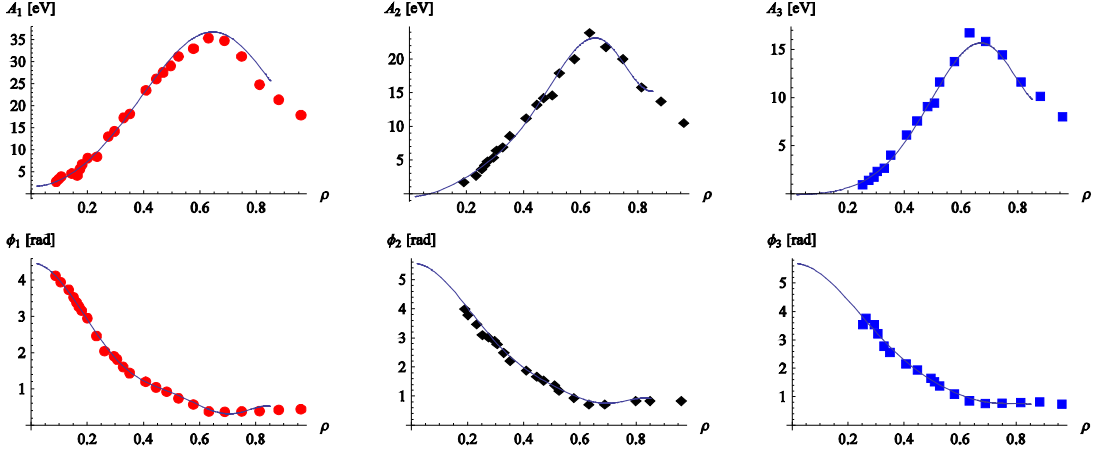

**Fig. 3**. Amplitudes (upper row) and phases (lower row) of measured electron temperature signal in AUG heat modulation experiments, adapted from Mantica et al. (2006). From left to right the fundamental frequency and its two next harmonics are reported (symbols), together with examples of interpolating curves used in Eq. (3.6).

The results are shown in Fig. (4), together with the best fit guessed in Mantica et al. (2006) within the framework of a Critical Gradient Model (CGM) plus a convective term adjusted to optimize the matching with the data. The reconstruction of coefficients using Eq. (3.6) is stopped short of the source location, at $\rho = 0.65$. The distinguishing feature of this result is the good overlapping between estimates of $\chi$ and (a bit more marginally) $V$ at different harmonics. This consistency check is an evidence that the simplest version of the convection-diffusion model given by Eq. (3.3) is relevant and that a single equation for temperature perturbations can model the physics for this particular experiment.

At the same time, Fig. (4) strikingly highlights the caveat concerning the issue of the multiplicity of allowed solutions, particularly severe using transport codes: the CGM solution is perfectly consistent with data, despite this, it is definitely not consistent with the MA one that has reliable error bars.

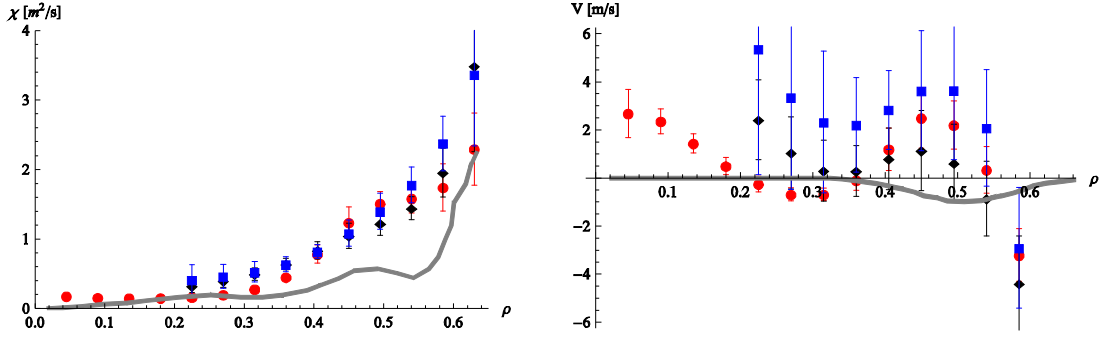

**Fig. 4.** Heat conductivity (left plot) and convective term (right plot). In accordance with common usage, the letter $\chi$ is used instead of $D$. Symbols are the results of Eq. (3.6) as computed from the data in Fig. (3). The same symbols and colours are used throughout the two figures. Results from the second and third harmonic are restricted to the $\rho > 0.2$ region, since phases are unreliable below this threshold (look at fig. 3). The solid curve is the best fit from Mantica et al. (2006).

For an example of failure of the CDM, we turn again to the JET experiment analyzed in Fig. (2), and namely consider the ICRH modulation, whose results appear in panels (a,b) of that figure. We recall that the duty cycle of the modulation is such that two harmonics are dominant: the fundamental one at $\nu_1 = 14.5$ Hz, and the first odd overtone $\nu_3 = 3\times\nu_1$.

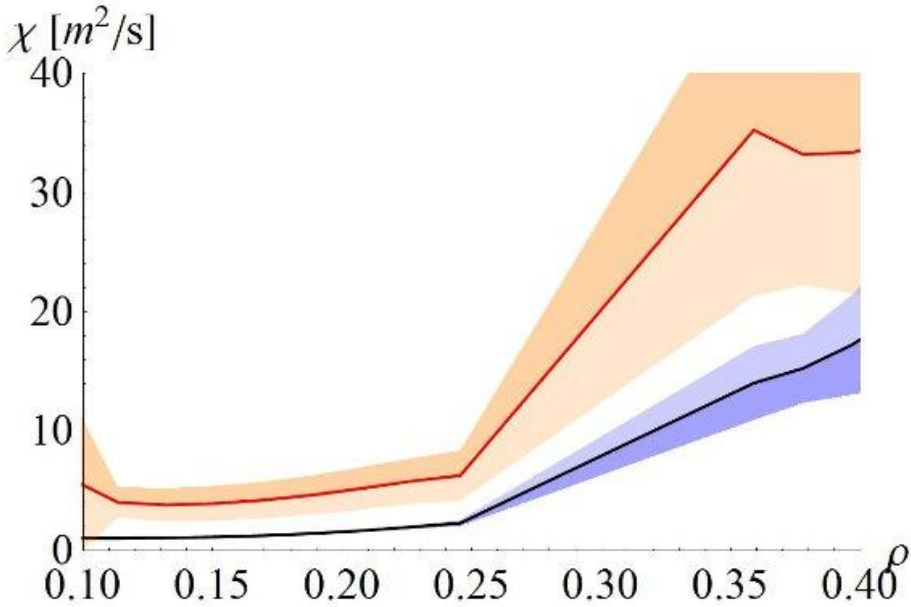

**Fig. 5.** Heat diffusivity reconstructed from the data in Fig. (2a,b) using the Matricial method. Blue curve refers to the dominant harmonic, red curve to the overtone. Colored areas display the confidence intervals. Adapted from (Sattin and Escande 2013a).

Fig. (5) displays heat diffusivity, computed applying the MA independently on the two sets of data; we do not show the pinch term since the associated error bars are too large to allow for any definite conclusion. There is a clear-cut disagreement between the two independent estimates in Fig. (5). This proves the failure of Eq. (3.3) for this example, as claimed in Section 2.

## 4 Other approaches to reconstruction of transport coefficients

*4.1 Pulsed experiments*

The MA algorithm has been known for about twenty years, and might have enjoyed more consideration. However, it is designed for periodic signals only, which rules out the important class of experiments with pulsed sources. In principle, Eq. (3.6) can potentially be adapted to studying these experiments as well: it is sufficient to go directly from Eq. (3.3) to Eq. (3.6) without the intermediate step of the Fourier transform (3.4), and retaining the time derivative *df/dt* in the r.h.s. of (3.6) which is now a real-valued equation. Accordingly, two or more instances of it with different data sets $(f, df/dr, df/dt)$ evaluated at different times need to be written down in order to assemble a linear system. The drawback is that the results might dependent on the chosen sets of data. In order to minimize this ambiguity it is necessary to make the system over determined by using several data sets and then to infer (*D,V*) via minimization algorithms, for instance by minimizing the functional

$$F = \sum_{i,j} \left[ D\frac{\partial f}{\partial r} - Vf + r^{-1}\int_0^r z\left(S(z,t) - \frac{\partial}{\partial t}f(z,t)\right)dz \right]^2_{r=r_i, t=t_j} \quad (4.1)$$

where measurements are taken at points $\{r_i\}$ and at times $\{t_j\}$. Experimental time derivatives can be approximated by linear interpolation between successive measurements, or with more refined schemes, such as splines, if enough time measurements are available.

Moret et al. (1992) adopted an approach from systems theory in terms of poles of the plasma response function, which in principle can encompass any kind of stimulus, both pulsed and modulated ones, and resembles the present one. Instead of using directly the time derivatives, this algorithm uses the Laplace transform in Eq. (3.3). Since the practical implementation of the algorithm requires the evaluation of all quantities at several frequencies in order to compute numerically their Laplace transforms, it is doubtful whether any practical advantage exists over the direct algorithm with time derivatives (4.1).

*4.2 Transport code approach using Genetic Algorithms to improve search strategies*

Due to its large flexibility the most widespread approach to the analysis of perturbative experiments is provided by transport codes. These codes use the direct numerical integration in time and space of transport equations like (3.3) for an a priori defined set of profiles of *(V,D)*, and they choose the profile providing the best agreement of the numerical *f(r)* with the measured one. Some popular codes either standalone or as part of a suite, are ASTRA (Pereverzev and Yushmanov 2002), JETTO (Cenacchi and Taroni 1988) or CRONOS (Basiuk et al. 2003); see also Gentle et al. (1987) and Baker et al. (1998). In this section we introduce an appealing approach to alleviate the search of a global optimum solution: Genetic Algorithms (GAs). These algorithms aim at solving optimization problems. First they can span in parallel considerable fractions of the whole solution manifold, thereby reducing the risk of exhibiting a local optimum. Second, they dynamically expand the solution space itself: starting from a reduced set of candidate solutions, new elements may progressively be added as requested by the data themselves, without the need for a direct intervention of the researcher.

More precisely, GAs are numerical search tools aiming at finding the global optimum of a given real objective function of one or more real variables, possibly subject to various linear or non linear constraints. Introductions to GA are found, e.g., in (Holland 1992, Koza 1992, Mitchell 1998). A population of individuals ("chromosomes"), each one representing a point within the solution space, collectively evolves towards better solutions through a random process of (1) parents' selection, (2) reproduction, (3) mutation and (4) substitution in the population. The parents' selection step determines the coupling among individuals which then yield an offspring in the reproduction phase; reproduction itself allows for the exchange of already existing genes whereas mutation introduces new genetic material; the substitution defines the individuals for the next-generation population. Such methods are commonly known as metaheuristics as they make few or no assumptions about the problem being optimized and can search very large spaces of candidate solutions. This way of proceeding enables in principle to arrive efficiently at optimal or near-optimal solutions; however, metaheuristics *do not guarantee* an optimal solution is ever found. The target of the optimization procedure is to find the point in the model parameters space which gives rise to the optimum of a properly assigned objective function, called fitness. Each chromosome is then made of as many ''genes'' (sub-strings) as the number of parameters to be determined. Each gene codes a possible value of a parameter so that each chromosome represents a possible solution to the problem: for each such chromosome, encoding a set of values of the parameters, the model is run and its outputs are used to compute the chromosome fitness. The population of chromosomes, then, evolves in successive generations ruled by the four fundamental operations. One example of applying GA to transport reconstruction is provided by (Giacobbo et al. 2002).

Differential Evolution (DE) (Storn and Price 1997) is a specific subset of the broader space of GAs, with the following restrictions: (I) The genotype is some form of a real-valued vector; (II) The mutation / crossover operations make use of the difference between two or more vectors in the population to create a new vector.

DE performs well for multidimensional real-valued functions because the vectors can be considered to form a "cloud" that explores the solution space quite effectively, automatically adjusting to span preferentially the regions where the function to be maximized takes its largest values. DE does not use the gradient of the function being optimized, which means DE does not require for the optimization problem to be differentiable as is required by classic optimization methods such as gradient descent and quasi-Newton methods. DE can therefore also be used in optimization problems that are not even continuous, are noisy, change over time, etc … . DE optimizes a problem by maintaining a population of candidate solutions and creating new candidate solutions by combining existing ones according to its simple formulae, and then keeping whichever candidate solution has the best score or fitness on the optimization problem at hand. In this way the optimization problem is treated as a black box that merely provides a measure of quality given a candidate solution, and the gradient is therefore not needed. Furthermore, this feature is particularly interesting for the present context, since it implies that GA's naturally evolve from an initial space of candidate solutions, that can possibly be quite restricted, and tend to sample the available space in a clever manner, since regions of solutions space which score bad values of the fitness functions are penalized during the stages (1-2) and do impact only marginally over the computational burden. Hence, the volume of solution space is varied self-consistently by the algorithm itself on the basis of the dataset. One recent very interesting example exploiting this feature in fusion physics is provided by the study by Murari et al. (2013).

In our case the function to optimize is the distance of a PDE solution from a set of measures coming from the phenomenon described by the PDE, using trial parameters. These parameters are the "genetic code" of one individual. The algorithm works on few hundreds of individuals.

By their very own nature, GA's tend to be computationally demanding. Luckily, the present algorithm is very parallelizable: on a multicore machine, every core operates on a vector independently of the others. In particular, for the present study we considered the option of using GPU's, a hardware which is becoming more and more popular due to its very appealing performance/cost ratio. In our implementation of the DE algorithm, the vectors are placed in the memory of the board, we need 2 kernels (i.e. procedures running on the device) for the computation of the DE operations: one kernel implements the combined mutation-crossover part and another the selection. The latter operation is accomplished after the fitness function is computed for each vector. This is done using a third kernel that solves the PDE using a Crank-Nicholson algorithm.

In order to test the algorithm, we implemented it first using C# language for rapid prototyping of the computational part and of the graphic user interface. Then we implemented a version on the GPU in CUDA, a set of directives (a subset of a programming language) specifically written for controlling GPUs. More details about the algorithms are summarized in Appendix C.

*4.2.1 Test cases with synthetic data*

We consider the usual convective-diffusive equation in cylindrical geometry, with radius $0 < r < 1$, which we rewrite here for convenience of the reader

$$\frac{\partial \xi}{\partial t} = \frac{1}{r}\frac{\partial}{\partial r}\left[r\left(\chi\frac{\partial \xi}{\partial r} - V\xi\right)\right] + S \qquad (4.2)$$

Eq. (4.2) is used to produce synthetic data. Coefficients $(\chi, V)$ are given in advance of the computation and are reproduced in Figs. (6,7). Initial and boundary conditions are set as $\xi = 0$ at $t = 0$, $\xi(r=1) = 0$, $d\xi/dr(r=0) = 0$. Source $S$ is taken as a narrow Gaussian centred near the edge and fading away very quickly. By solving Eq. (4.2) we produce some datasets $\{\xi(r_i, t_j)\}$: an instance is shown as dots in Fig. (6), and each colour stands for a different time $t_j$. Times $t_j$ are chosen in the time interval where the source has disappeared.

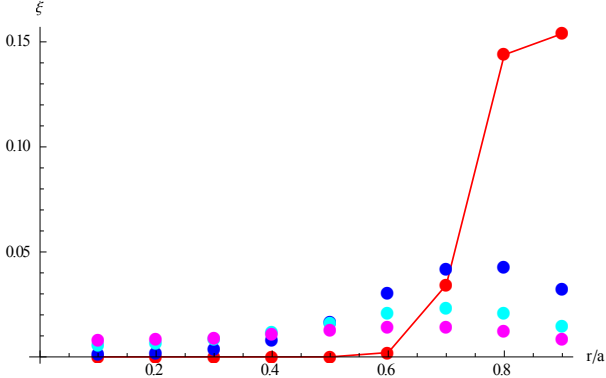

**Fig. 6.** Examples of synthetic datasets produced using Eq. (4.2). The half-life of the source is 0.5 (arbitrary units). Data are sampled at $t$ = 5,10,20,40.

For the modelling stage, the first set $\{\xi(r_i,t_1)\}$ (red dots joined by a curve in Fig. 6; for numerical convenience in the code it is interpolated by a Gaussian curve) is used as initial condition. In detail, the Cranck-Nicholson algorithm solves a variant of Eq. (4.2):

$$\frac{\partial \xi}{\partial t} = \nabla \cdot \left( \chi_{mod} \nabla \xi - V_{mod} \xi \right) \quad (4.3)$$

using $\{\xi(r_i,t_1)\}$ as initial condition. For the sake of simplicity we did not implement explicitly the source $S$, although it may be included easily. The coefficients $\chi_{mod}$, $V_{mod}$ are the unknown parameters the optimization algorithm attempts to assess by matching against raw data (blue, cyan and magenta dots in Fig. 6). Presently, we parameterize them in terms of linear combinations of radial basis functions (polynomials):

$$\chi_{mod}, V_{mod} = \sum c_i r^i, i = 0,...,N, \quad (4.4)$$

with the constraint — due to symmetry considerations — that only even powers are used for $\chi$, and only odd ones for $V$. The number of terms retained in the power series expansion is a knob left to the modeller: a number of sensitivity tests can be devised in order to assess the optimal number for each case.

We repeated the execution of the algorithm using all the combinations of number of terms for the two series between $N=1$ and a fixed maximum ($N=7$). This is a brute force, but exhaustive approach. We preferred it since runs were carried out on a sufficiently powerful hardware.

The robustness of the DE algorithm against external noise is estimated through a Monte Carlo scheme: the data $\{\xi(r_i,t_j)\}$ $j$=2,…, 4, are polluted adding a 5% error independently to each datum and the corresponding coefficients evaluated. Several independent replica simulations are performed, which allowed getting estimates of the variances. Notice that the DE is a stochastic algorithm, thus some amount of variance is expected even without adding explicitly noise.

The figure (6) below shows the very satisfactory agreement obtained using $N$=4 coefficients (hence, a total of 8 free parameters). Note however, that it must be understood as an overly optimal situation, since it emulates exactly the coefficients ($\chi,V$) employed in Eq. (4.2) to produce the synthetic data: fourth-order polynomials. It is however, encouraging, that the algorithm be able to recover very effectively the true solution.

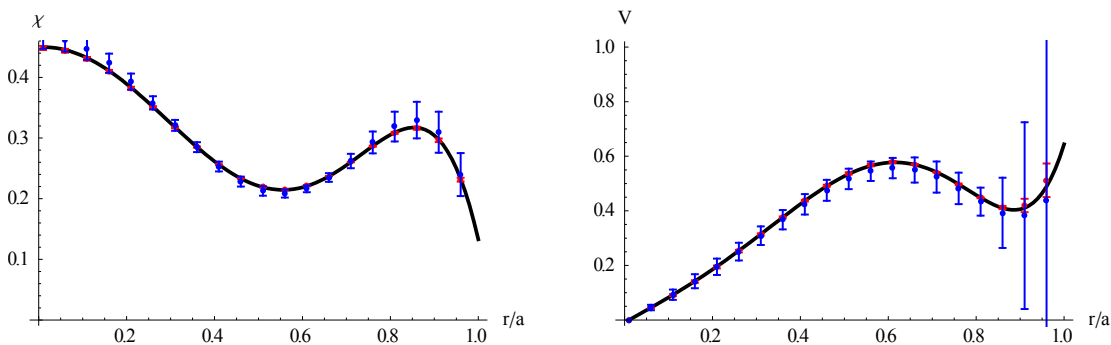

**Fig. 7.** Left plot, conductivity $\chi$, right plot convection $V$. Black curves are the true coefficients; red dots, their estimates by the DE using the raw data in Fig. (6) without added noise (error bars are of the same size as the dots); blue dots with error bars, the same with a noise level of 5% of the signal added. Four basis functions, $N$ = 4, were used for both $\chi$ and $V$.

Had we allowed a smaller maximum number of parameters, the quality of the interpolation would have been sensitively worse. This example highlights the necessity of using sufficiently flexible model basis function, and therefore of working in relatively high-dimension parameter spaces.

A more stringent test is illustrated in Fig. (8). There, the true transport coefficients ($D,V$) contain exponential functions, thus particularly difficult to be matched by any polynomial combinations, which provides a further example of the caveat previously mentioned, namely it is not warranted that even the best approximation reconstructed from transport codes is actually close to the true solution, if the space of potential solutions, chosen beforehand of any calculations, is not wide enough.

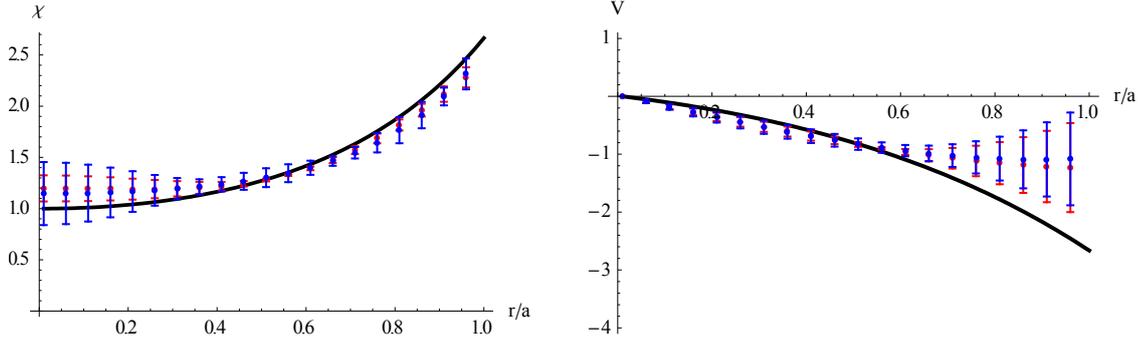

**Fig**. **8**. Another example of reconstruction of transport coefficients, like Fig. (7). True transport coefficients are displayed as thick black curves; points with error bars are the result of the inversion algorithm.

Two relevant facts now are: (i) the quality of the fit degrades somewhat, as expected. However, even accounting for the error bars, there is strong disagreement between the true coefficients and the reconstructed ones only upon *V* for *r* > 0.7 — where the data are scarce. (ii) Interestingly, the error bars turn out almost the same polluting or not the data with noise. This means that — in this case at least — the largest source of error is provided by the stochasticity intrinsic to the DE.

We conclude this section by recalling once again that, even though GA appear very promising tools, they cannot override the limitations to the reconstruction of transport highlighted by the Matricial Approach, which are intrinsic to the data set. In other words, near to a source, where the linear system (3.6) yields one vanishingly small eigenvalue, and therefore strong instability to small experimental errors, one cannot expect the GA to converge towards just a single sharply resolved optimum.

**5 Issues with the reconstructions of transport codes**

The problem we are referring to can be cast as: to which extent the approaches to the reconstruction of the transport coefficients like those presented in section 4.2 may be considered reliable. This topic was already addressed somehow implicitly in Section 3, and more explicitly in Section 4. Indeed, when using transport codes, the proof of the accuracy of the guessed solution is replaced with that of the data reconstruction. This is not an equivalent statement, since reconstructions closely matching the experimental data can be obtained with sets of transport coefficients very different among them, and far from the true ones. One instance of the severity of this issue was given in fig. (4): the estimates of (*D*,*V*) given using the Matricial Approach (symbols) and the Critical Gradient Model (curves) are definitely different: in particular, opposite signs are predicted for the pinch term *V*. Nonetheless, both sets of coefficients, when inserted back into the transport equation (3.4), produce data that match the experimentally measured ones, within the error bars. For the Matricial Algorithm this is true by construction, since it provides an optimum solution for any minimization algorithm (see Sattin *et al.* 2012). For the CGM, one may refer to the original reference Mantica et al. (2006).

The purpose of this section is to provide a few further instances of this same issue, in order to better highlight how much generic it can be in common situations.

*Test case #1*

We start by resorting again to the paradigmatic case of the JET discharge 55809: we showed in Section 2 (Fig. 2) that the plasma response, in this case, is not driven by transport. Section 3 (Fig. 5) backs up this claim: the transport reconstruction of the coefficients using the Matricial Algorithm yields inconsistent estimates between the different harmonics of the ICRH modulation. Despite this, we show here that one (or possibly several) couples of reasonable (*D,V*) profiles can be guessed: (I) the profiles shown in the first row of Fig. (9) are patterned after those produced in Fig. (5). As such, they are expected to yield a good matching of ICRH modulation data, but not a perfect one, since Fig. (5) showed that it is not possible to model simultaneously data from both harmonics using just one set of coefficients. (II) In the second row, the matching with cold pulse data is very good at the outermost chord ($r = 0.3$); the agreement is clearly much worse at the innermost chord ($r = 0.11$)—which is expected since, the mechanism driving this dynamics *is not* transport. However, as argued by del-Castillo-Negrete et al. (2008), the qualifying feature is the time of arrival of the perturbation, which is flagged by the change in the slope of the time trace, and which takes place at about 3 ms both for the experiment and the simulation.

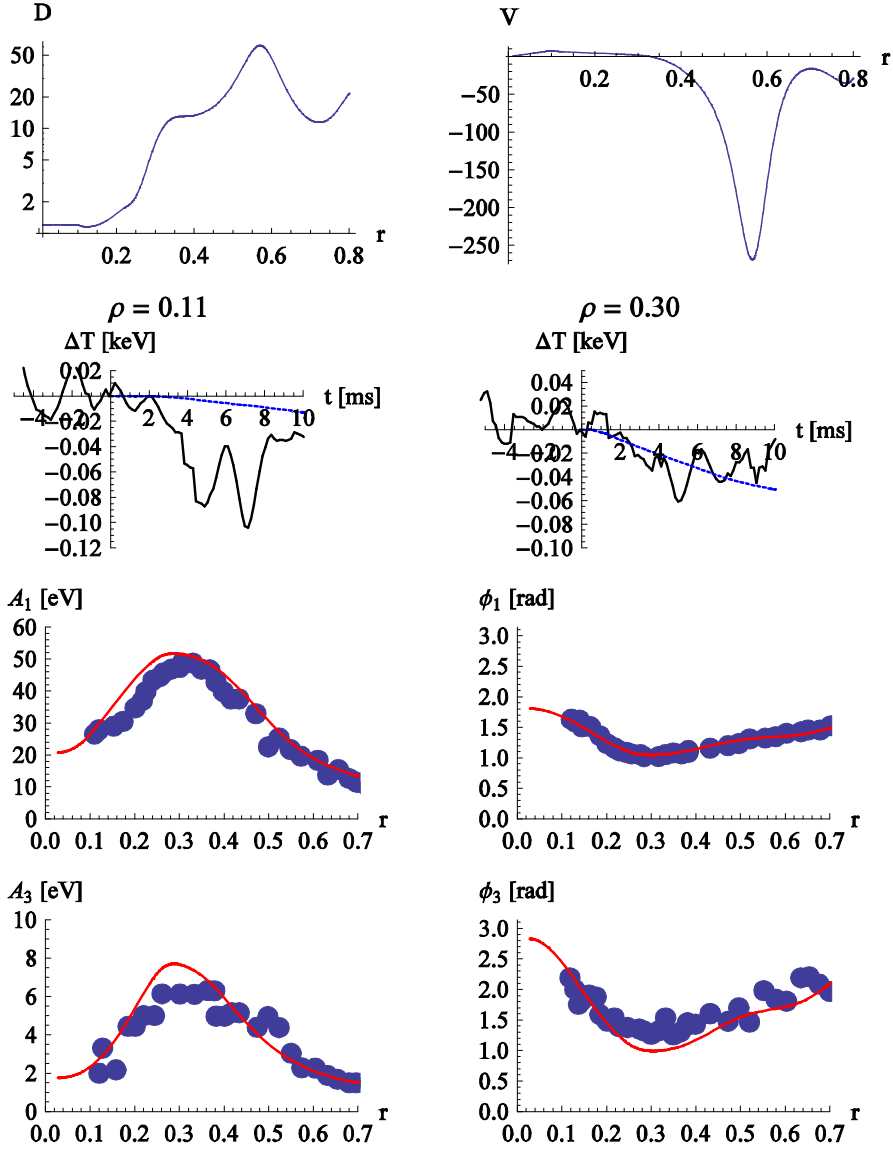

**Fig. 9.** First row: diffusivity (*D*) and convection (*V*) used in the simulation. Second row, the experimental (solid curves are the same as in Fig. 2) and numerical (dotted curves) time traces for the cold pulse propagation. The third and fourth rows contain the same ICRH modulation data (dots) as shown in Fig. (2), solid curves are the numerical results.

*Test case #2*

As a second example we turn to density transport experiments carried out at LHD by Tanaka et al. (2006) via modulated gas puffing and summarized in Fig. (10). The measurement of electron density fluctuations is carried out through a 13 channel laser FIR interferometer (but only ten channels were used). Extracting radial profiles from integrated line-of-sight measurements is an inverse problem in itself, and therefore subject to quite large uncertainties. Finally, there is potentially a considerable larger impact from the coupled heat perturbation. The analysis technique used in the original reference was integrating Eq. (3.4) with trial functions for (*D,V*). No sophisticated spatial trends were considered, but some piece-wise constant ($D, V/r$). The particle source was estimated using the DEGAS code (Heifetz et al. 1982). The optimization was performed by minimizing the sum of squares between integrated measurements and reconstructed ones. In this case, too, the final result yields figures that overlap the experimental ones within the error bars (see fig. 10 of Tanaka et al. 2006).

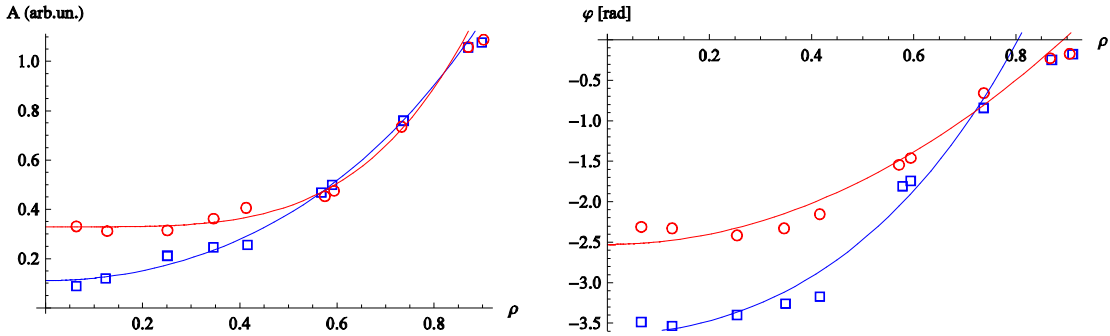

**Fig. 10**. Symbols, measured modulated amplitude (left figure) and phase (right figure), adapted from Fig. (7) of Tanaka et al. (2006). Red circles and blue squares have the same meaning as in Tanaka et al. and refer to data taken in a discharge with 5.2 (respectively 1.0) MW of NBI heating power. Solid curve are examples of interpolating functions.

Our modelling was carried out just like in the previous example. The source term was neglected since, according to the DEGAS result, it drops down by about three orders of magnitude with respect to its edge value for $\rho < 0.7$. Thus, the code is able to recover the correct qualitative trend for both coefficients (Fig. 11), but there is a quantitative disagreement which cannot definitely be recovered within the confidence intervals. In this case, we did not have available redundant measurements for carrying out consistency checks, therefore it is not possible to assess the role and importance of coupled transport channels.

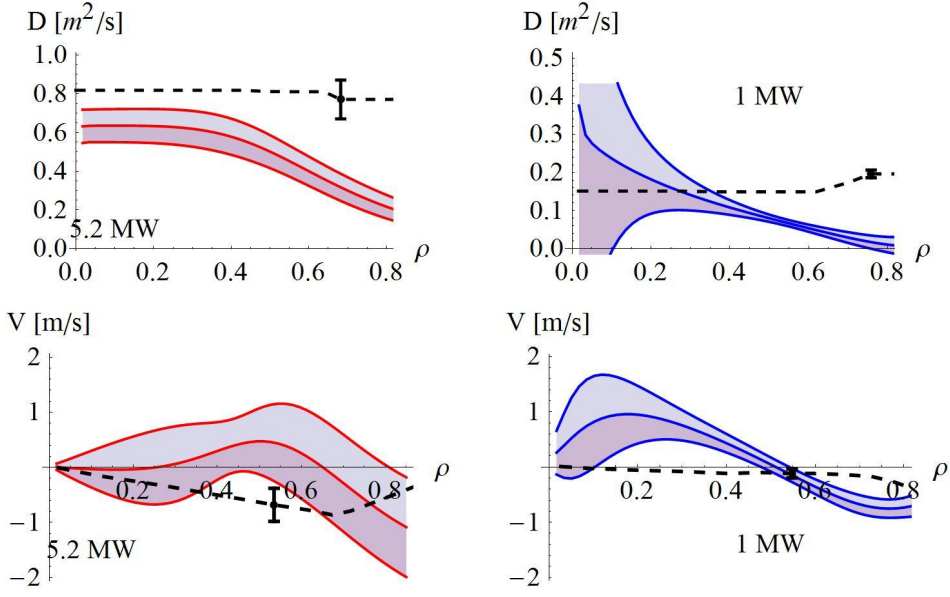

**Fig. 11**. Upper row, diffusivity; lower row, convection. Red curves refer to the high-power discharge, the blue ones to the low-power one. The coloured areas envelop the confidence intervals. The dashed curves are the best fit estimated in Tanaka et al. (2006) by matching the modulated part of the signal.

*Test case #3*

A third example is a synthetic one. We produce a set of synthetic data by integrating Eq. (3.3) using known transport coefficients and sources. In order to check a different scenario, we consider the case of a pulsed source and we collect data during the decay phase after the source was switched off. Transport coefficients are

$$D = 1, V = r/2 - r^3 + 4r^5. \tag{5.1}$$

The system starts with zero initial density, $f = 0$. At time $t = 0$ a source term $S$ (a very narrow gaussian) arises near the edge, and vanishes suddenly at $t = 0.01$. Eq. (3.3) is integrated numerically with standard boundary conditions ($df/dr = 0$ at $r = 0$, $f = 0$ at $r = 1$), and is then sub-sampled in a way to mimic a hypothetical experiment: we sample over a grid with steps $dr = 1/15$ and $dt = 0.0175$. Samples were taken during the "free decay" phase of the signal, well after the source was switched off. In order to model the presence of some noise, we add to each "measurement" $f_{ij}$ some random quantity taken from a normal distribution with 1% amplitude of the clean signal. The discrete set $\{f_{ij}\}$ is interpolated with smooth cubic splines, along both the radial and the temporal direction.

On this occasion we cannot employ the Matricial Approach in the version outlined in section 3. We attempt then to infer (*D,V*) from this set of data: by (i) minimization of the functional (4.1). (ii) by considering two instances of Eq. (3.3), at two different times ($t_1,t_2$) taken during the decay phase — after the switching off of the source. These two equations define a linear algebraic system whose solution, performed in the spirit of the Matricial Approach, yields (*D,V*). We checked that, with the sampling adopted, final results turn out independent of the choice of ($t_1,t_2$) within a wide enough interval ($t_2 - t_1$). Minimization of *F* (4.1) was performed postulating (*D,V*) to be of the form $D = \exp(d_0 + d_2 r^2), V = v_1 r + v_3 r^3$. The particular functional form for *D* was chosen in order to insure positiveness, whereas we purposely constrained *V* to an analytical expression that is insufficiently generic for tracking the true convective term, in order to investigate the consequences of fitting with too narrow a solution space. Finally, we tested several global search algorithms: Differential Annealing and Differential Evolution, as well as local ones as Conjugate Gradients, all of them built into the Mathematica software. The results turn out independent of the algorithm used.

Results are shown in Fig. (12), both in the absence and presence of added noise. The striking conclusions reached by inspection of Fig. (12) is that the addition of just a little amount of noise is sufficient in this case to completely deceive the optimization algorithm. Conversely, the Matricial algorithm turns out more robust: both in the presence or absence of noise, it fits well the true solution.

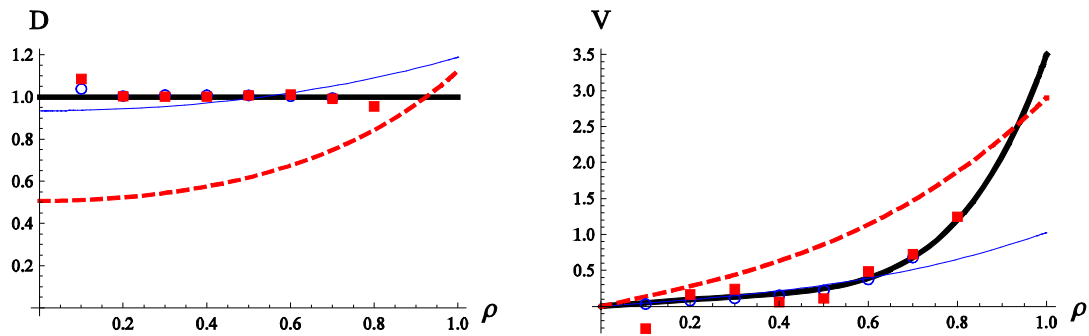

**Fig. 12**. Diffusivity (on the left) and convection (on the right) as reconstructed from the Matricial algorithm (symbols) and the optimization one (curves). In both plots, black curves are the true coefficients. Blue circles and blue lines stand for the result of the modeling, respectively using the Matricial algorithm and minimizing Eq. (4.1) in the absence of noise. The red squares and dashed curves are the corresponding results in the presence of noise at 1% amplitude of the signal.

Figure (12) is instructive since it contains both factors that may limit the performance of reconstruction algorithms, as highlighted earlier: in the absence of noise the blue curve does not fit perfectly the correct solution since this is prevented by the class of analytical functions used for the fit. By expanding sufficiently this class, it is easy to reach an excellent matching. In order to understand the behavior of equation (3.2) in the presence of noise, we first rewrite it by setting $S = 0$ and postulating that $f$ is self-affine in time, $f = g(r)\exp(-t/\tau)$, which is a fairly good approximation:

$$-\frac{g}{\tau} = \frac{1}{r}\frac{\partial}{\partial r}\left[r\left(D\frac{\partial g}{\partial r} - Vg\right)\right] \qquad (5.2)$$

Equation (5.2) takes the form of an effective steady state transport equation. At this stage, transport coefficients are no longer uniquely defined: to any solution $(D_1, V_1)$, we can add another one $(D_2, V_2)$ such that its driven flux is zero: $\Gamma_2 = -D_2\frac{dg}{dr} + V_2 g \equiv 0$. This is exactly what happens in our case. We define

$$\begin{cases} \Gamma = -D\frac{dy_{mes}}{dr} + Vy_{mes} \\ \hat{\Gamma} = -D_{reg}\frac{dy_{mes}}{dr} + V_{reg}y_{mes} \end{cases} \qquad (5.3)$$

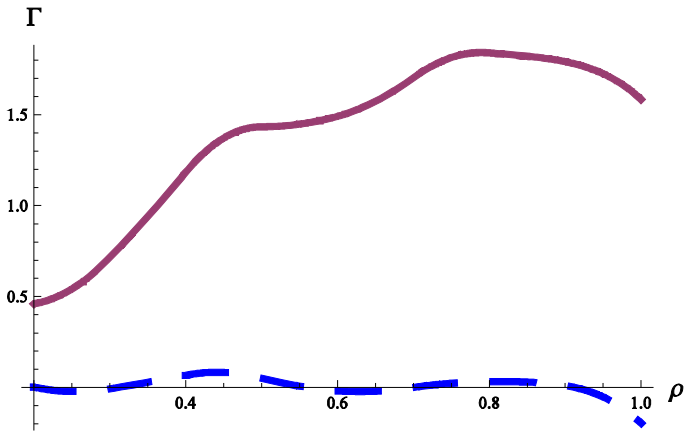

**Fig. 13**. The thick brown line is the flux $\Gamma$ driven at $t = 0.18$ and defined in the first of Eqns. (5.3), the dashed blue curve is the difference $\Gamma - \hat{\Gamma}$, defined in the same equation, at the same time.

In (5.3) $y_{mes}$ is the "measured" signal: that is, the signal produced synthetically with the addition of noise; $(D,V)$ are the true transport coefficients, whereas $(D_{reg}, V_{reg})$ are the coefficients estimated by (4.1) — the dashed curves in Fig. (12). Figure (13) displays $\Gamma$ (thick curve) and $\Gamma - \hat{\Gamma}$ (dashed curve) at a given time. Despite the transport coefficients are widely different, the driven flux is almost the same: the minimization of Eq. (4.1) falls onto a solution that differs by an additional contribution ($D_{reg}$-$D$, $V_{reg}$-$V$) to the true coefficients.

In summary, the cases shown above are understood to provide a wide selection of failing inferences of transport coefficients despite obtaining good reconstruction of experimental data. There are several possible reasons for these failures: the transport model itself is inadequate in case #1; the space of candidate profiles is chosen too narrow in case #2 and #3, where furthermore the pollution by numerical noise is seen to be able to shift the found solution away from its true value.

## 6 Summary and comments

In Section 2 and Appendix A we argue that the origin for some fast plasma dynamics, in a first instance attributed to transport, is MHD in reality. Indeed, in this case the inadequacy of the transport paradigm itself is clearly reflected in the impossibility of reconstructing reasonable guesses for the transport coefficients: this was clearly highlighted in Section 3 (Fig. 5) as well as 5 (Fig. 9).

In most cases, however, the situation is not so clear-cut: reconstruction of transport coefficients by transport codes may easily yield apparently satisfactory answers, in terms of agreement with experimental data, without necessarily be close to the true solution. An algorithm able to yield the exact solution: the MA, is recalled in section 3. The MA represents, whenever applicable, an appealing solving strategy: it is exact (as long transport scenarios as in Eq. 3.4 are involved), provides explicit and easy ways of assessing the confidence intervals, and is computationally light.

On the other hand, shortcomings of the transport codes can be attributed from the one hand to their difficulty in converging toward the true solution, and from the other hand to the lack of a rigorous criterion for assessing whether the true solution has been reached. Several instances of these pitfalls are evidenced in Sections 3,4, and extensively in Section 5.

In order to improve over the limitations of transport codes outlined in the previous points, we suggest that the implementation of Genetic Algorithms, outlined in Section 4 and Appendix B, is a promising tool, since they involve global search strategies and, furthermore, are able to dynamically and automatically evolve the full space of solutions.

On the other hand, one has to consider further the possibility that the transport paradigm is still valid, but no satisfactory conclusions can be reached because the model does not account properly for all the quantities actually involved. This is the case where several transport channels are critically coupled together, but the modeling retains just few of them. Instances are shown in Appendix C. Thus a further caveat of this work is stressing the importance of retaining as wide as possible transport models. This holds for the analysis stage as well as the preparatory one: that is, experiments should be designed having in mind the need of monitoring as many transport channels as possible.

Finally, we add some comments concerning the findings of Sections 2 and 3: It appears from these evidences that magnetized plasmas may exhibit a wide set of dynamics, driven by different physics mechanisms, likely coupled among them. A very recent study draws a conclusion that can be considered as affine to ours: Ettoumi et al (2013) argue that, during sawteeths, reconnection of magnetic field — a resistive MHD phenomenon — is incomplete, leading to a fraction of stochastic magnetic field lines; the fast disruptive relaxations might be caused by the ensuing fast parallel transport of matter and heat. Thus, in this case, too, MHD produces an actual transport of plasma. Also in the paper by Nicolas et al (2014) the transport of impurities during sawteeth is satisfactorily modeled treating them as passive scalar quantities advected by MHD flows. The mutual interaction between micro-turbulence and large-scale MHD structures is a topic that is being investigated self-consistently (Ishizawa and Nakajima 2007, Muraglia et al 2009, Hornsby et al 2010, Ichiguchi and Carreras 2011, Li and Kishimoto 2012, Schlutt et al 2013). The present work, that uses only linear models, cannot contemplate the interaction between the two physics. Rather it studies the effective response of the plasma and infers which of the two mechanisms is prevailing. It provides an understanding which might be improved by further dedicated experiments. In particular, in order to check the coupling of different transport channels, it would be advisable to perform modulation experiments were more than one actuator-response couple would be measured carefully. Then an extension of the Matricial Approach might check the quality of the reconstructions provided by a series of Convection-Diffusion models of increasing complexity.

## Acknowledgements

Y. Camenen is acknowledged for making us aware of some references and for discussions. Discussions with A. Salmi and P. Mantica were useful for developing some of the ideas presented here. M.-C. Firpo provided one reference. F.S. and D.E. wish to thank the organizers of the 8$^{th}$ Workshop on Fusion Data Processing, Validation and Analysis where a version of this work was presented and for some illuminating talks and discussions. G.U. acknowledges Ionoco Ltd. for providing the GTX Titan card used for CUDA code development. This work was supported by the European Communities under the contract of Association Euratom/ENEA. The views and opinions expressed herein do not necessarily reflect those of the European Commission.

## Appendix A. Computation of the $v^{(0)}$ flow and derivation of Eqns (2.3,2.4)

*Part A.1*

The relevant equations for this analysis are, out of the whole set (2.1), just the force balance equation and the Faraday-Ohm one. Since we are interested into equilibrium conditions, we can dispose of the time derivatives. From the other side, we cannot neglect any longer resistive effects. Therefore we include a finite constant resistivity η in Ohm's law. Using normalizations introduced in Section 2, the equations write

$$u\frac{dP}{dr} + \frac{5}{3}\frac{P}{r}\frac{d}{dr}(ru) = \frac{1}{r}\frac{d}{dr}\left(r\chi_p\frac{dP}{dr}\right) + S \tag{A1}$$

$$\beta\frac{dP}{dr} + B_z\frac{dB_z}{dr} + \frac{B_\theta}{r}\frac{d}{dr}(rB_\theta) \tag{A2}$$

$$uB_z - \eta\frac{dB_z}{dr} = 0 \tag{A3}$$

$$-uB_\theta + \frac{\eta}{r}\frac{d}{dr}(rB_\theta) = \text{cost} \tag{A4}$$

We now consider an expansion in powers of the small parameter β. Typical figures for tokamaks show that $(Rq)^{-2} \approx O(\beta)$. Given this condition, the leading terms in the solution of the force balance equation (A2) are $(B_\theta, B_z) = (r/(Rq), 1)$. Thus, we write $B_z = 1 + \beta b_z$, $u = u_0 + \beta u_1$, and insert them into (A3). The corresponding zeroth and first order terms yield

$$\begin{cases} u_0 = 0 \\ u_1 = \eta\frac{db_z}{dr} \end{cases}$$

Hence, the order of magnitude of the equilibrium flow, in normalized units, is β×η. The Spitzer resistivity for a plasma with temperature of order $10^{2 \div 3}$ eV is $\eta_{SPITZER} \approx O(10^{-6 \div -8})\Omega m$. It is normalized, in our units, by division with the quantity $\eta_0 = u_A \mu_0 a B_0^{-2} \approx O(10)$. Furthermore $\beta \approx O(10^{-2})$: the combined product is by some orders of magnitude smaller than typical figures recovered for the perturbed velocity **v** encountered in Section 2, $O(10^{-7})$. A more quantitative estimate can be obtained by expressing the pressure and the poloidal field in terms of $u_1$, by means of (A2, A4): eventually, (A1) becomes a differential equation for $u_1$ alone.

*Part A.2*

Let us now consider Eqns. (2.1), normalized and disregarding $\mathbf{v}^{(0)}$:

$$\frac{\partial \rho}{\partial t} + \nabla \cdot \mathbf{v} = S_n + \chi_n \nabla^2 \rho \tag{A5}$$

$$0 = \mathbf{J}^{(0)} \times \mathbf{B} + \mathbf{J} \times \mathbf{B}^{(0)} - \nabla p \tag{A6}$$

$$\frac{\partial p}{\partial t} + 2\beta \mathbf{v} \cdot \nabla T^{(0)} + \beta \frac{10}{3} T^{(0)} \frac{\partial \rho}{\partial t} = S_p + \chi_p \nabla^2 p \tag{A7}$$

$$\frac{\partial \mathbf{B}}{\partial t} = \nabla \times (\mathbf{v} \times \mathbf{B}^{(0)}) \tag{A8}$$

Let us differentiate (A6) with respect to time, and (A7) with respect to space; replace inside them time derivatives of ρ, **B**, using (A5, A8) and neglecting diffusive terms.

$$\begin{aligned}\frac{\partial \nabla p}{\partial t} &= (\nabla \times \mathbf{B}^{(0)}) \times [\nabla \times (\mathbf{v} \times \mathbf{B}^{(0)})] + \{\nabla \times [\nabla \times (\mathbf{v} \times \mathbf{B}^{(0)})]\} \times \mathbf{B}^{(0)} \\ \frac{\partial \nabla p}{\partial t} &= -\nabla \left[ 2\beta \mathbf{v} \cdot \nabla T^{(0)} + \frac{10}{3}\beta T^{(0)} (S_n - \nabla \cdot \mathbf{v}) - S_p \right]\end{aligned} \tag{A9}$$

Equating these two equations leads to a second-order ordinary differential equation for **v**. Notice that retaining the diffusivities would imply the appearance of third-order spatial derivatives $d^3 v / dr^3$ in (A9): these terms are expected to be smaller than the others, since we do not look for wildly fluctuating profiles of the flow. A posteriori, this is checked in our simulations, since—except near the sources—*v* is almost a linear function of the radius.

**Appendix B. Some instances of the importance of the coupled channels in transport modeling**

We use as reference the RFX-mod Reversed Field Pinch (Martin et al. 2011). One promising operational scenario of this machine is the SHAx state, featuring electron temperature internal transport barriers. It is of current interest to understand whether internal barriers do exist for particle transport, too. In RFX-mod, matter can be delivered in the core only through deep pellet fuelling. In most cases, the perturbation caused by the pellet is so strong that it spoils the magnetic geometry, but in the two cases examined here, the magnetic configuration was seen to survive for some milliseconds.

The particle density profile was monitored with high-time resolution through inversion of the interferometer signal. All other transport channels — and most notably the electron temperature one — are not accounted for: only quite recently a high-sampling-rate diagnostics for monitoring electron temperature profiles was installed in RFX-mod (Franz et al. 2013).

In the two figures (14,15) we show the dynamics of the particle density profile during the decay phase, after the pellet has been totally ablated. Postulating that the particle density equation is decoupled from the rest of the transport, it reduces to

$$\frac{\partial n}{\partial t} = \frac{1}{\rho}\frac{\partial}{\partial \rho}\left[\rho\left(D\frac{\partial n}{\partial \rho} - V n\right)\right] + S_n \qquad (B1)$$

The whole analysis turns out to be conveniently performed in terms of just one helical flux coordinate as radial coordinate. The geometry is taken as cylindrical for simplicity, the small error introduced does not affect qualitatively the conclusions. The source term $S_n$ stands for the residual particle source due to the neutrals coming from the walls: it is extremely small in the core.

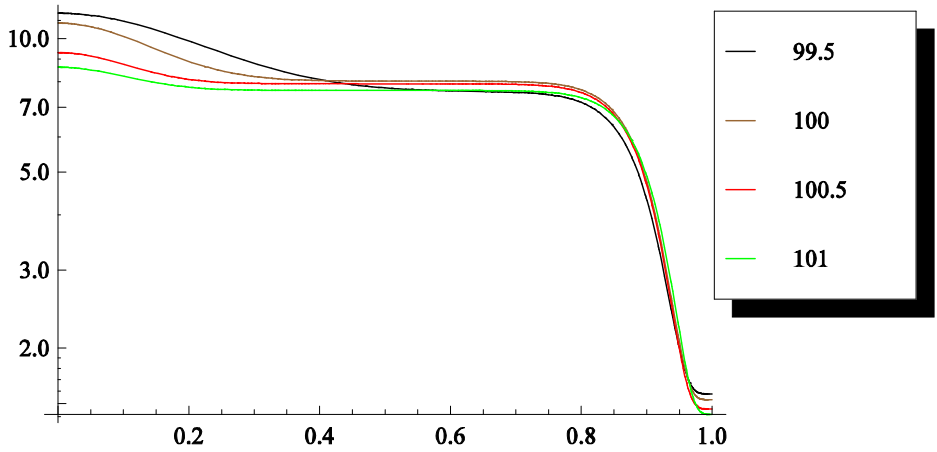

**Fig. 14**. Inverted density profiles for discharge 23945 between 99.5 and 101 ms.

It is straightforward to realize that the trend shown in Fig. (14) cannot be modelled by Eq. (B1): the role of the diffusivity is to spread any initially localized structure, whereas in this case there is a shrinking of the central bulge. Mathematically it could only be described in terms of a negative diffusivity, $D < 0$.

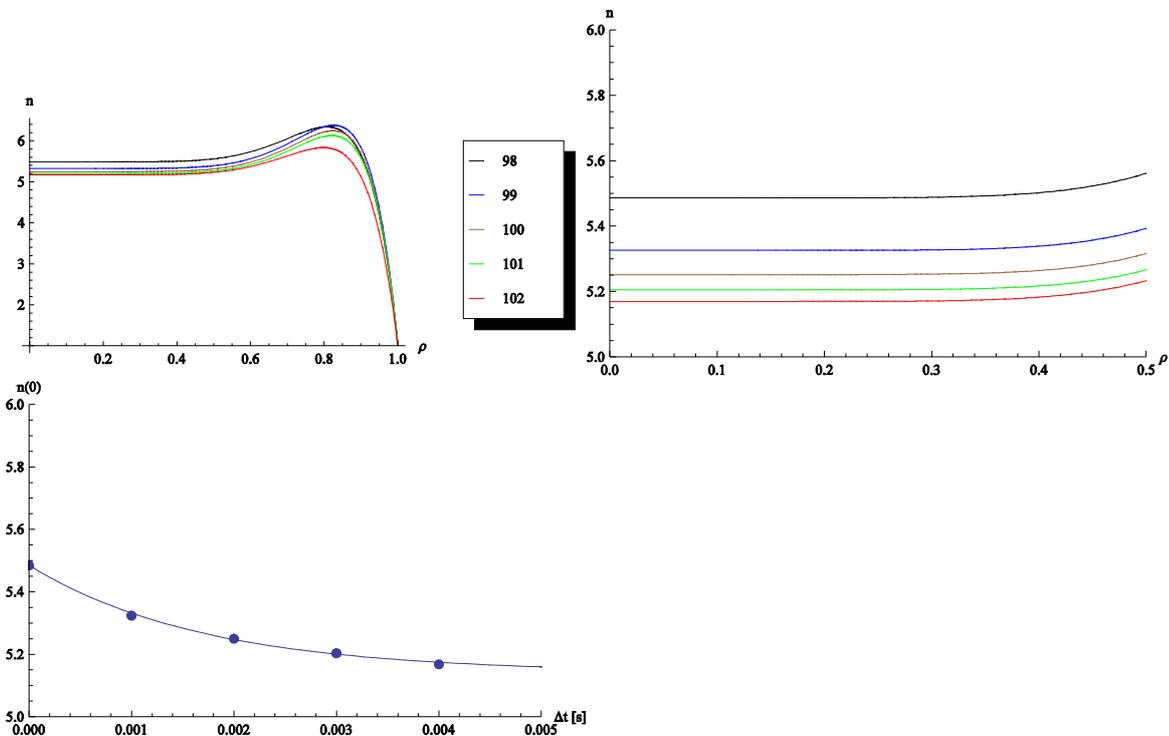

**Fig. 15**. Top row, inverted density profiles at several times from 98 ms to 102 ms for RFX-mod discharge 24953 and, on the right, zoom of the same figure. Bottom plot, dots stand for the value of the central density at the several times. The solid curve is the fit (B2). Density units are $10^{19}$ m$^{-3}$.

The analysis is a little trickier for the case shown in Fig. (15). In this case, limiting to the very central region, say $\rho < 0.2$, the density is flat therein, $dn/d\rho = 0$. Hence, Eq. (B1) can be solved exactly by postulating

$$\begin{cases} n = n_\infty + \delta n \times \exp(-t/\tau) \\ V = V_0 \times \rho \\ S_n = \text{constant} \end{cases} \quad (B2)$$

When fitted this expression to the data from Fig. (15), the result is $n_\infty = 5.15\ m^{-3}$, $\delta n = 0.34\ m^{-3}$, $\tau = 1.7\ ms$, $S_n = 3 \times 10^{22}\ m^{-3} s^{-1}$. The value of $\tau$ is consistent with present estimates for particle confinement time in RFX-mod, whereas $S_n$ found this way is completely inconsistent with the numerical simulation of neutral gas transport inside the plasma performed with Monte Carlo code NENE (Lorenzini et al. 2006), that predicts a core source term about three order of magnitude smaller, $\approx O(10^{19})\ m^{-3} s^{-1}$.

Unlike the discussion in Section 2, here there are no particular reasons to question the convective-diffusive paradigm *per se*. Rather, it is its specific implementation (B1) which is likely too simplistic. For instance, we stressed in Section 3 that the neglect of coupled transport channels may translate into spurious source/sinks terms. It is plausible that temperature be affected into the very core, hence those spurious terms should appear there, too, which is exactly the region where our modeling appears to fail most.

**Appendix C. Details of the Differential Evolution algorithm**

After the random initialization of the population, a cycle consisting of three main operations (crossover, mutation and selection) is repeated until an exit condition is reached.

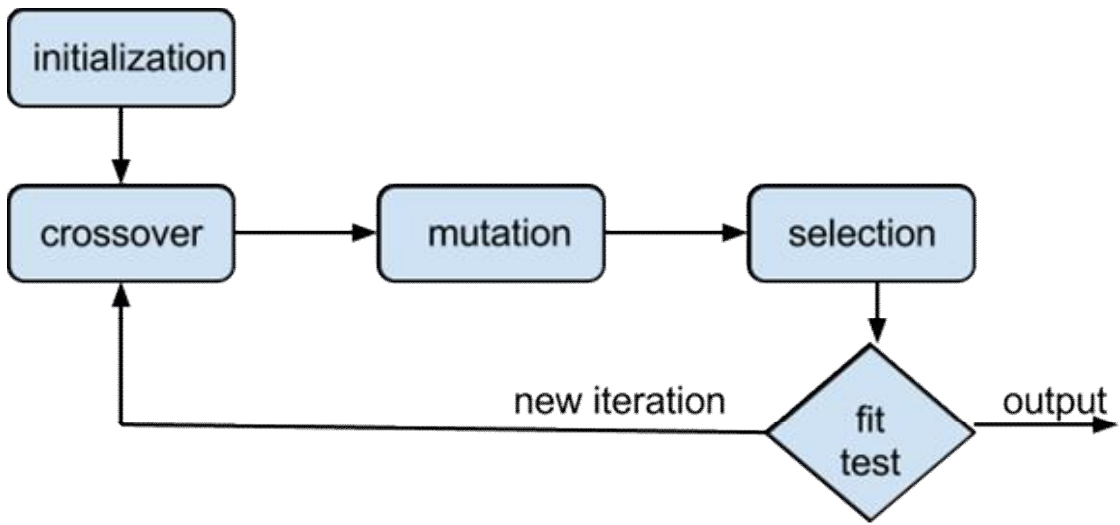

**Fig. 16**. Flowchart of DE algorithm

Definitions:

Individual: array of real numbers

Population: set of individuals

*NP*: size of the population

*n*: dimension of the arrays

*G*: generation index, i.e. the number of times the algorithm has been evaluated.

*X*: generic individual

$X_{i,G} = (x_{1,i,G}, \ldots, x_{n,i,G})$: i-th individual of the population after *G* iterations of the algorithm $i \in \{1, \ldots, NP\}$

$x_{j,i,G}$: j-th component of the array $X_{i,G}$ $j \in \{1, \ldots, n\}$

*V*: individual generated by the mutation

*U*: individual generated by the crossover

Initialization: each component of the *NP* individuals is generated randomly in the corresponding range [min, max] given by the user for that component:

$x_{j,i,0} := \text{RNG}(\min_j, \max_j) \quad j \in \{1,\ldots,n\}, i \in \{1,\ldots,NP\}$

Mutation: The mutation operation of DE applies the vector differentials between the existing population members in order to determining both the degree and direction of perturbation applied to the individual subject of the mutation operation. The mutation process at each generation begins by randomly selecting five individuals in the population ($X$). The i-th perturbed individual, $V_{i,G+1} = (v_{1,i,G+1}, \ldots, v_{n,i,G+1})$, is then generated on the chosen individuals using one of the following formulas (Wang et al. 2011):

1) $V_{i,G+1} = X_{i,G} + F \times (X_{r1,G} - X_{r2,G})$

2) $V_{i,G+1} = X_{best,G} + F \times (X_{r1,G} - X_{r2,G})$

3) $V_{i,G+1} = X_{i,G} + F \times (X_{best,G} - X_{r3,G}) + F \times (X_{r1,G} - X_{r2,G})$

4) $V_{i,G+1} = X_{best,G} + F \times (X_{r4,G} - X_{r3,G}) + F \times (X_{r1,G} - X_{r2,G})$

5) $V_{i,G+1} = X_{r5,G} + F \times (X_{r4,G} - X_{r3,G}) + F \times (X_{r1,G} - X_{r2,G})$

The operation is repeated for each individual $i$, where $i \in \{1,\ldots,NP\}$; $r_1, r_2, r_3, r_4, r_5 \in \{1,\ldots,NP\}$ are indexes randomly selected such that $r_1 \neq r_2 \neq r_3 \neq r_4 \neq r_5 \neq i$; $F$ is the control parameter, its precise value is not relevant as long as it stays within the interval [0, 2]. $X_{best,G}$ is the individual with the best fitness in the generation $G$. The whole population is divided into five groups, for each group a different formula is used.

Unlike other GA, the mutation is stochastic (the $X$ are randomly chosen) but not random since the new values are computed using given formulas.

Crossover: The perturbed individual, $V_{i,G+1}$, and the current population member, $X_{i,G}$, are then subject to the crossover operation, that generates the population of candidates $U_{i,G+1} = (u_{1,i,G+1}, \ldots, u_{n,i,G+1})$, as follows:

$u_{j,i,G+1} = v_{j,i,G+1}$   if $r < Cr$ or $j = k$. $Cr \in [0, 1]$, the crossover rate, is chosen arbitrarily.

$u_{j,i,G+1} = x_{j,i,G+1}$   otherwise

The operation is repeated for each component $j$ and each individual $i$, where: $j \in \{1,\ldots,n\}$, $i \in \{1,\ldots,NP\}$, $k \in \{1,\ldots,n\}$, $k$ is a random component's index chosen once for each individual $i$ in order to change at least one component.

Selection: The population for the next generation is selected from the individuals in current population and the candidates generated by the crossover according to the following rule:

$X_{i,G+1} = U_{i,G+1}$ if $f(U_{i,G}) < f(X_{i,G})$

$X_{i,G+1} = X_{i,G}$ otherwise

where $f$ is the objective function.

A fitness function $f(U_{i,G})$ is computed for each candidate $U_i$, and it is compared with the fitness $f(X_{i,G})$ of its counterpart $X_i$ in the current population. The one with the lower value will survive from the tournament selection and will become the individual $i$ of the population of the next generation $G+1$. As a result, all the individuals of the next generation are better than their counterparts in the current generation.

Exit condition: The cycle is repeated until the lowest fitness value (i.e. the fitness of the *best* individual) falls below a given tolerance or if a maximum number of iterations is reached.